\documentclass[onecolumn]{aastex631}
\usepackage{lineno}

\begin{document}

\def\msol{\hbox{$\hbox{M}_\odot$}}
\def\lsol{\hbox{$\hbox{L}_\odot$}}
\def\kms{km s$^{-1}$}
\newcommand{\pram}{P_\mathrm{ram}}

\title{The Population of the Galactic Center Filaments: Position Angle Distribution\\
       Reveal a Degree-scale Collimated Outflow from Sgr A* along the Galactic Plane
}

\author[0000-0001-8403-8548]{F. Yusef-Zadeh} 
\affiliation{Dept Physics and Astronomy, CIERA, Northwestern University, 2145 Sheridan Road, Evanston , IL 60207, USA
(zadeh@northwestern.edu)}

\author[0000-0001-8403-8548]{R. G. Arendt} 
\affiliation{Code 665, NASA/GSFC, 8800 Greenbelt Road, Greenbelt, MD 20771, USA}
\affiliation{UMBC/CRESST 2 (Richard.G.Arendt@nasa.gov)}

\author[0000-0001-8403-8548]{M. Wardle} 
\affiliation{Research Centre for Astronomy, Astrophysics
and Astrophotonics, Macquarie University, Sydney NSW 2109, Australia, (mark.wardle@mq.edu.au)}

\author[0000-0001-8403-8548]{I. Heywood} 
\affiliation{Astrophysics, Department of Physics, University of Oxford, Keble Road, Oxford, OX1 3RH, UK}
\affiliation{Department of Physics and Electronics, Rhodes University, PO Box 94, Makhanda, 6140, South Africa}
\affiliation{Department of Physics and Electronics, Rhodes University, PO Box 94, Makhanda, 6140, South Africa 
(ian.heywood@physics.ox.ac.uk)}

\begin{abstract}
We have examined the distribution of the position angle (PA) of the Galactic center filaments with lengths $L > 66''$ and $ < 66''$ as well as their length
distribution as a function of PA. We find bimodal PA distributions of the filaments, long  and short  populations of radio filaments. Our PA study shows the
evidence for a distinct population of short filaments with PA close to the Galactic plane. Mainly thermal short radio filaments ($<66''$) have PAs concentrated
close to the Galactic plane within $60^\circ < \rm PA <120^\circ$. Remarkably, the short filament PAs are radial with respect to the Galactic center at $l
<0^\circ$, and extend in the direction toward Sgr A*. On a smaller scale, the prominent Sgr E HII complex G358.7-0.0 provides a vivid example of the nearly radial
distribution of short filaments. The bimodal PA distribution suggests different origin for two distinct filament populations. We argue that alignment of the short
filament population results from the ram pressure of a degree-scale outflow from Sgr A* that exceeds the internal filament pressure, and aligns them along the
Galactic plane. The ram pressure is estimated to be 2$\times10^6\,$ cm$^{-3}\,$ K at a distance of 300pc, requiring biconical mass outflow rate $10^{-4}$ \msol\,
yr$^{-1}$ with an opening angle of $\sim40^\circ$. This outflow aligns not only the magnetized filaments along the Galactic plane but also accelerates thermal
material associated with embedded or partially embedded clouds. This places an estimate of $\sim$6 Myr as the age of the outflow.
\end{abstract}


\section{Introduction} 

The prototype magnetized radio filaments in the Galactic center (GC) were discovered close to 40 years 
ago \citep{zadeh84}. Since then, VLA, Nobeyama 45m-telescope, Mologlo and ATCA   observations have shown linearly polarized 
synchrotron emission 
tracing nucleus-wide 
cosmic ray activity throughout the inner few hundred parsecs of the Galaxy 
\citep{liszt85,bally89,gray91,sofue92,lang99,reich00,larosa05,zadeh04,nord04,law08a,pound18,staguhn19,arendt19}. 
{\it Chandra}, {\it XMM} and {\it NuSTAR} have  also detected X-ray emission from a handful of nonthermal radio filaments.
 \citep{sakano03,lu03,zadeh05,zhang14,zhang20,zadeh21}

High temperature molecular clouds with large turbulent linewidths also characterize this region,   the Central Molecular 
Zone (CMZ). Furthermore, H$_3^+$ absorption line observations show that high cosmic ray ionization rates permeate the CMZ 
at levels a thousand times that of the solar neighborhood \citep{oka20}. These observations indicate that cosmic ray 
pressure in the CMZ 
is significant when compared to the interstellar gas pressure of the Galactic center.

Recent MeerKAT observations have provided a remarkable mosaic of the inner few degrees of the Galactic center (GC) with 
exquisite details at $4''-6''$ spatial resolution \citep{heywood19,heywood22}.
Statistical properties of the population 
of GC filaments such as  morphology, mean spectral index, equipartition magnetic field, filament spacing and their 
associations with stellar candidates have recently been studied in detail in a series of papers 
\citep{zadeh22a,zadeh22b,zadeh22c}. 
VLA observations showed the bright and 
long filaments emitting non-thermal emission whereas MeerKAT data showed a great deal of  fainter filaments and 
filamentary structures that  are shorter.  
 MeerKAT has also surveyed a large number of galaxy clusters, some of which show 
exceptional isolated filamentary structures in the intracluster medium (ICM). These filaments 
appear to be linked to radio jets, 
tails and lobes of radio galaxies
\citep{shimwell16,ramatsoku20,condon21,rudnick22}. 
For the first time, a population of magnetized ICM filaments, with an underlying 
physical mechanism similar to that of GC filament population, is identified in external galaxies \citep{zadeh22d}.

With the new and sensitive MeerKAT images of the Galactic center, we continue examining the physical properties of the GC 
filaments, focusing on their position angle (PA). This physical characteristic could potentially provide additional 
insight into the origin of the enigmatic GC filaments. Previous VLA observations of the brightest GC filaments showed 
that they are 
mainly oriented perpendicular to the Galactic plane (\cite{zadeh04} and references therein). Using high sensitivity of 
MeerKAT data, we examine statistical properties of the PA of the GC filament population.  We present  evidence 
that short filaments located on the negative Galactic longitudes, have PAs that run mainly within $\pm20^\circ$ of the 
Galactic plane. An extension of the filament PAs show radial distribution, pointing in the direction toward Sgr A* at the 
Galactic center.  The orientation of a large number of filaments within a few hundred pc of Sgr A* is similar to those 
 found in radio continuum and recombination line images of
the inner few arcminutes of  Sgr A* \citep{royster19,zadeh20}. 
In addition, ALMA observations of the Sgr E molecular cloud shows parallel filamentary 
structure along the Galactic plane \citep{wallace22}. 
Here we consider a picture in which a  collimated  outflow, most likely arising from 
Sgr A*, is responsible for 
distorting and aligning  the shape of molecular, thermal and nonthermal features  along the Galactic plane. 


\section{Data Reduction}
\label{sec:data}

Details of the MeerKAT observations can be found in \cite{heywood22}. Here we briefly describe the 
observations and data processing. The final image is constructed from 20 MeerKAT pointings for a total of 144 hours 
on-source. The L-band (856$-$1712~MHz) system was used, with the correlator configured to deliver 4,096 frequency 
channels. The 1.28 GHz mosaic covers the inner $\sim$3.5$^{\circ}$~$\times$~$\sim$2.5$^{\circ}$ ($l$~$\times$~$b$) of the 
Galactic center. 
Spectral index ($\alpha$) measurements (where $I_\nu \propto \nu^\alpha)$) have used 16 sub-bands between 856 and 1712 MHz. The 
in-band spectral index values are measured simultaneously centered at 1.28 GHz \citep{heywood22}.
 
To enhance the visibility of the filaments, the mosaic image is filtered
using a difference of Gaussians to smooth noise and remove 
large scale backgrounds, as described in detail in \cite{zadeh22a}. 
Numerous fainter filaments are uncovered and stand out in the 
filtered image. To confirm the reality of 
faint filaments in the filtered image, we have found their counterparts in the unfiltered image. 
There are a large number of twisted short features associated with HII regions and nonthermal filaments. 
Our analysis includes the long  and short filaments, both of which  consist  of a mixture of   thermal and nonthermal features.
 The lengths ($L$) and mean PAs of the filaments are measured by application of the procedure $LOOPTRACING\_{AUTO}$ 
\citep{aschwanden10}.  A detailed account of this procedure will be given elsewhere.

\section{Results}

The 4$''$-resolution MeerKAT image of the Galactic center is shown in Figure 1a. A filtered image of the same mosaic 
image is shown in Fig. 1b with a resolution of $\sim 6.4''$. SNRs and foreground HII complexes are drawn with red circles 
on Figures 1a,b.  Our analysis excludes  these features that would otherwise  contaminate our analysis of individual 
filaments 
in the Galactic center.

\subsection{Filament length as function of the PA}

Figure 2 shows three different representations of color-coded identified filaments with different PA ranges
with  respect to Galactic North (PA=$0^\circ$),  
 traced by 
different color tables. Figure 2a shows  the full range of filament PAs between $0^\circ$ and $180^\circ$ (or 
equivalently, $-90\arcdeg$ to $+90\arcdeg$)  
with different colors. Color tables are used in order to distinguish between the filaments with positive (red) negative 
(blue)  PAs. The distribution suggests that the filaments oriented perpendicular to the Galactic plane tend to 
be long. On the other hand, filaments with PAs running parallel to the Galactic plane are short. Figure 
2b shows 
the distribution of short filaments with PAs being preferred along the Galactic plane 60$^\circ<$~PA~$<~120^\circ$. This 
distribution is more obvious at negative longitudes where short filaments with PA$\sim120^\circ$ appear blue and 
point 
toward the Galactic center. 
Figure 2c displays the filaments with $-60^\circ <$ PA $<60^\circ$ indicating that vertical 
filaments are dominated by long filaments at positive and negative latitudes. An extension of the long filaments toward 
the Galactic plane do not converge toward a single origin 
such as Sgr A* but to the  inner degree of the Galactic center.

\subsection{Long filaments PA}

Another way to characterize the distribution of filament PAs is by restricting their lengths rather than their PAs, as 
was described above. 
There is generally a continuum of  filament lengths, and  
66$''$  was chosen as a lower limit that eliminated most filaments associated with 
known thermal sources such as  HII regions. 
An even cleaner selection at 132$''$  was discussed, but the  sample size starts to 
become small \citep{zadeh22a}.
This convenient division  was selected using 60  pixels  with a pixel size of 1.1$''$ in the original MeerKAT image with a resolution  
of 4$''$ \citep{zadeh22a}.

Figure 3a shows a histogram of long filaments with $L>66''$ as a function of PA, emphasizing 
long nonthermal filaments,  and largely excluding thermal features.
A Gaussian fit to this histogram peaks at -3.2$^\circ$ with 1$\sigma$ error of 12.9$^\circ$. 
Figure 3b shows a histogram of long filaments with 
restricted  PAs ranging between $-60^\circ$ and $60^\circ$. We note that long filaments 
have a negative spectral indices 
$\alpha$ the spectral index 
consistent with being nonthermal. There are also a number of long filaments with  $\alpha\sim-0.15$. These also are most 
likely associated with a network of filaments associated with the Radio Arc near $l\sim0.2$, which are 
known to have a flat spectral index (\cite{zadeh22a} and references therein).

Another view of the PA distribution of the filaments is presented by plotting them as a function of Galactic longitude. 
The PA distribution of long   filaments as a function of Galactic longitude is displayed  in Figure 4a whereas the 
same distribution only for short  filaments is shown  on Figure 4b.  Each black  
dot represents the PA of a filament. If the filament PAs were randomly distributed, there would not be any concentration.  
 We note a high concentration of  PAs in two clusters, one  near   PA$\sim20^\circ$, 
$l\sim0.2^\circ$ and the other  near  PA$\sim-30^\circ$,  $l\sim-0.25^\circ$.  A higher concentration of 
filaments 
at $l>0^\circ$ is due to a larger number of filaments associated with the radio Arc near $l\sim0.2^\circ$. 
An additional concentration of short filaments is noted  near $l\sim 0.4^\circ$ with PAs along the Galactic plane,  
as 
demonstrated  in Figure 4b. 

We also note that similar to 
Figure 2c, the two concentrations in Figure 4a, show  that the PA of the long filaments are 
tilted away from 
the Galactic North. We note 
that 
at $l<0^\circ$ the filament PAs of the concentrations are more widely distributed than those at $l>0^\circ$. 
Figure 4a shows that within the range $0.5^\circ < l < -0.75^\circ$, long filaments generally are found with 
$PA \sim 0^\circ$, indicating a vertical orientation.
For $l > 0.5^\circ$, the long filaments have random orientations. For $l\sim -0.75^\circ$, long filaments are loosely 
clustered 
around $PA = \pm 90^\circ$.



\subsection{Short filaments PA}

One of the most striking aspects of our PA study of the GC filaments is the evidence for a distinct population  of 
filaments with  
PA close to the Galactic plane. 
This distribution is orthogonal to the PA distribution of long filaments that are 
aligned close to the Galactic North-South orientation. 
Figure 5a shows the PA distribution of short  ($L<66''$) 
filaments. 
The  short filament PAs are  modulated by two peaks near 70$^\circ$ and $-10^\circ$. The 
filaments in the  peak at PAs$\sim70^\circ$ with a width of $\sim40^\circ$ are distributed mainly  within 
$\sim\pm20^\circ$ of the the Galactic plane whereas the filaments that peak close to PAs$\sim-10^\circ$ are oriented 
closer to the 
Galactic North-South direction,  having a distribution similar to that of the long filaments. 
In some 
cases the filaments can only be 
reliably traced over partial segments of the filament due to limited  S/N and confusion.


Another characteristic that distinguishes short from long filaments is their spectral index. 
Figure 5b shows a histogram of the spectral indices of short filaments $L<66''$,  restricted to angles  within 
$\sim30^\circ$ of the Galactic plane ($60^\circ < PA < 120^\circ$).  
The majority  of short filaments have spectral indices consistent 
with thermal emission, though we can not exclude that they could be nonthermal filaments similar to the 
filaments of the radio Arc near $\sim0.2^\circ$.
This histogram also indicates that  some  short filaments are nonthermal with 
a steep spectrum. 

The distribution of PAs  is also examined as a function of the Galactic Position Angle (GPA) defined as 
$\mathrm{arctan} (l/b)$. $0^\circ< GPA <90^\circ$ corresponds to the NE  quadrant of the GC, and 
subsequent $90^\circ$ intervals are the SE, SW, and NW quadrants.
 The black  points represent 
the PA of short- and long-identified filaments in Figures 6a,b, respectively. The dark diagonal line shows the expected 
PA 
distribution along 
a radial direction following a PA which is equal to its GPA. Filaments along these lines are oriented radially with 
respect to the Galactic center.However, we note a concentration of points close to GPA$=270^\circ$ (close to the Galactic plane and W of Sgr 
A*) along the Galactic 
plane. 
There is a tendency that filament PAs in the third GPA$<270^\circ$ and fourth quadrant GPA$>270^\circ$ display 
horizontal components with PAs$\sim70^\circ$ and $\sim110^\circ$, respectively. This range of filament PAs
follow the diagonal line that points in the direction of Sgr A*. The width of the group of filaments following the 
diagonal line is about 20$^\circ$ above and below the Galactic plane.  
These trends indicate that filaments PAs both N and S of the Galactic plane tend to be slightly more normal (rather 
the parallel) to the Galactic plane than would be expected for a purely radial distribution of filaments.
Figure 6c schematically indicates that this trend could result from a strong radial trend (long magenta arrows). 
Deviations of filament PAs from radial direction could result from the direction of orbital motion of individual 
sources when superimposed with radial direction of the outflow.




\subsubsection{Short filaments PA in Sgr E}

A more detailed representation of the radial distribution of short filament PA pointing toward the Galactic center can be 
viewed in the spatial distribution of filaments PA in 
the prominent Sgr E HII complex G358.7-0.0,  providing  a 
vivid example of the nearly radial distribution of short filaments.
This star forming 
cloud 
contains a number of discrete HII regions which have been imaged   using the  VLA, GBT and MeerKAT 
\citep{gray93,gray94,cram96,zadeh04,law08b}. Figure 7a displays an unfiltered 20cm continuum MeerKAT image of the Sgr E 
discrete HII regions distributed  around the southern half of a 0.4$^\circ$ ring-shaped structure 
\citep{zadeh04}. The northern half 
of  Sgr E is dominated by extended diffuse emission.
Some  filamentary features in Sgr E  appear to be correlated with structure  in IR emission. 
There is  no correlation in 
structure between
traditional nonthermal radio filaments and IR emission. This is one  reason that we argued  that 
short filaments could be thermal.

Figure 7b is similar to 7a except that a filtered image more clearly  shows the  large number of filamentary features. The 
white contour 
shows the outer boundary of $^{13}$CO (2-1) emission from the Sgr E cloud with a mass of 3$\times10^5$ \msol 
\citep{anderson20}. The $^{13}$CO mean velocity is $-170$ \kms\, in the southeast, becoming more negative to --220 \kms\, 
toward northwest \citep{anderson20}. The identification of specific filaments with measured PAs are indicated
 by red lines in Figure 7c.  The histogram of 
of PAs of these filaments in Sgr E, as shown in Figure 7d,  indicates that filament 
PAs peak near PA$\sim92.3^\circ$ within $1\sigma = 39^\circ$, as the Gaussian fit (red) shows. 
The PA distribution  in Sgr~E is consistent with larger-scale distribution of filament PAs at negative Galactic longitudes, 
thus 
implying that the  same mechanism is responsible for the origin of the filaments on  
this  smaller, sub-degree scale. 

This pattern of radial structure close to the Galactic plane  
is seen on even small scale observations of some of the 
compact molecular clouds in the region. 
Recent ALMA observations of CO  associated with the Sgr E complex finds two molecular filaments running close 
to the Galactic plane within ALMA's 52$''$ field of view.
These molecular filaments have an aspect ratio of $\sim5$ to 1 
and show  alignment  
within 2$^\circ$  of  the Galactic plane \citep{wallace22} appear to align radially toward the Galactic center. 
The origin of this filamentary structure is suggested  to be due to stretching of 
the Sgr E cloud by the gravitational field of the Galactic bar potential.  
The PAs of the CO filaments resemble 
those found in our radio continuum study, so it is possible that they are produced by the mechanism responsible for 
aligning the radio filaments. 
MeerKAT radio image shows a short filament G358.712+0.027 with an extent of $\sim1'$ and 
a position angle 
of $106\arcdeg$ (a deviation of $15\arcdeg$ from a radial orientation), possibly associated with the molecular 
filament.  

Figures 7a,b also display  a broadly triangular dark 
feature bounded by the lines of H II regions on the SE and SW, and also the CO contours on the 
SW and N.  The  mean brightness of 47 $\mu$Jy beam$^{-1}$,  corresponding
 to brightness temperature of $\sim2.2^\circ$K,  
coincident with contours of molecular CO in the 
southern half of the ring-shaped structure.   To examine if this dark feature is real,  we 
examined 
radio continuum image of the Galactic center at 8 GHz observed with the GBT \citep{law08b}.  
The dark feature is clearly seen toward the Sgr E cloud and the estimated brightness temperature toward the cloud, 
$\sim9.7\times10^{-3}$ Jy beam$^{-1}$, is a factor of $\sim$2 lower than the background diffuse emission. 
The deficit of $\sim 10$\,mJy per $88^{\prime\prime}$ GBT beam is equivalent to an emission measure drop of 
$n_e^2L_c\approx 560\,\mathrm{cm^{-6}\,pc}$ for a line of sight through the radio-dark cloud.  Adopting a pathlength $L_c 
= 14\,pc$ through the cloud, equivalent to 0.1$^\circ$ at 8\,kpc, we find an equivalent electron density deficit 
$n_e\approx 6\,\mathrm{cm^{-3}}$.  Making the reasonable assumption that  the scale of external ionized medium has  
similar path length, it  provides a  rough estimate of the density of the external ionized medium surrounding 
the cloud.

The dark feature in the continuum MeerKAT and GBT images 
is  anti-correlated with the Sgr E CO cloud. 
This dark feature shows similar appearance  to many Galactic center molecular clouds that have been 
identified as  Radio Dark 
Clouds (RDCs) \citep{zadeh12b}. We interpret that the
dark feature in Sgr E  surrounded by compact HII regions is  
 produced by a deficiency in radio continuum emission from 
molecular clouds that are embedded
in a bath of UV radiation.  The deficit  of the continuum
emission from the volume occupied by molecular gas 
results in dark features that trace embedded molecular clouds.




 
\section{Discussion}

Our analysis of the Galactic center filaments PA distribution indicate  two different filament populations.  
The PAs of the long and short filaments tend to 
lie  in the directions perpendicular and 
parallel to the Galactic plane, respectively. The vertical and horizontal filaments are  characterized to 
have  
nonthermal and thermal properties \citep{zadeh22a}, respectively, based on their spectral indices. 
While there is clearly an overlap in their 
properties (e.g., PA, spectral index)  
a key question is  the origin of the bimodal PA distribution of Galactic center filaments.

\subsection{Vertical  Population of Galactic center filaments PA}

One population of filaments has orientations that are vertical to within $\sim\pm25^\circ$ of the Galactic North. The 
negative and positive PAs tend to be found in positive and negative Galactic longitudes, respectively. There is slightly 
reduced number of filaments at PAs$\sim0^\circ$, giving roughly a bimodal distribution in the histogram. We also note that 
vertical filaments dominated by long filaments $L>66''$ have nonthermal spectra. Considering global distribution of 
vertical filament PAs, there is no evidence that extension of 
all filaments PA converge toward Sgr A* or any other compact sources. 
This  suggests that the 
 creation  of roughly 
symmetric PA distribution is in 
the Galactic plane within a degree of Sgr A*.
 This is 
consistent with 
a trend noticed in the spectral 
index of nonthermal filaments as they become  steeper  with increasing absolute Galactic latitude \citep{zadeh22a}. 
If synchrotron aging is  responsible for the steepening  the spectrum, it  implies  that the vertical filaments 
are  powered  in the Galactic plane \citep{zadeh22a}. 

A scenario for the creation of  nonthermal radio filaments uses the high cosmic ray pressure in the Galactic 
center estimated 
from H$_3^+$ measurements \citep[e.g.][]{geballe99,oka05,oka20}. This extreme pressure drives a large-scale wind away from 
the Galactic plane creating the bipolar X-ray and radio emission \citep{heywood19,ponti19,zadeh19}. The non-thermal radio 
filaments result from the interaction of the large-scale wind and obstacles embedded within the flow creating the 
filaments by wrapping of the wind's magnetic field around the obstacles \citep{zadeh19}. 
However, there are alternate  models 
explaining the origin of the vertical filaments (see \citep{zadeh22a,sofue23} and references therein). This begs the 
question of the origin of the high cosmic-ray pressure which is usually attributed to the explosive event a few million 
years ago that created Fermi bubbles and the bipolar X-ray and radio emission \citep{heywood19,ponti19,zadeh19}.

\subsection{Radial   population of filaments PA in the Galactic plane}

The most interesting result of our study is the discovery of a new  population of short filaments' PAs that are found  
along the Galactic plane only at $l < 0^\circ$. 
There is  statistical evidence 
that short filaments PA point toward Sgr A* with 
the cone in which the filaments are found has FWHM$\sim20^\circ$, 
but the 
deviations of filament directions from radial is much 
broader,  over a large scale of 
$\sim300$  pc from Sgr A* at
$l < 0^\circ$. This  large-scale anisotropy in the filaments PA is interpreted in the context of a large-scale collimated 
outflow from 
Sgr A* along the plane. 
A wind- or jet-driven outflow from Sgr A* has been suggested 
to explain evidence for outflow in the plane of the Galaxy within few arcminutes  of  Sgr A*.

There are eight  different spectroscopic, polarimetric  and broad-band continuum measurements over a wide range of 
angular scales 
extending from $\sim0.02$  to $\sim$25 pc from Sgr A* have inferred  a jet-driven outflow 
along the Galactic plane (see \cite{zadeh12a,zadeh20} and references therein). 
In this picture the jet emerges perpendicular to the equatorial plane of the accretion flow and is 
aligned roughly E-W in Galactic coordinates.  
A schematic diagram in Figure 8 showing the interaction of the outflow from Sgr A* with filaments, ionized and molecular gas being 
aligned along roughly along the Galactic flow. 
For example,  blue- and red-shifted features in  radio recombination line observations on a scale 
within 14$''$ (0.56 pc)  of Sgr A* invoke 
an interpretation in terms of the interaction of a 
collimated outflow driven outflow  from Sgr A* along $\sim\pm30^\circ$ of the Galactic plane  
\citep{royster19}. 
In this picture, the blue-shifted arm of the jet emerges to the west  of Sgr A* making an angle 
$\sim45^\circ$ to the line of sight.


This small-scale geometry of the jet is consistent with our analysis of filaments PA if the two opposing cones 
of outflowing  material are  extrapolated to a scale of $\sim$300 pc.  
Thus, the blue-shifted component of the jet  projected  along negative longitudes and is 
tilted toward us with respect
 the line of sight
whereas the red-shifted component is tilted away from the line of sight projected toward the positive longitudes. 
However, 
there is  a  lack of strong radial filament PAs at positive longitudes in the direction away from the 
line of sight. This could be  explained by confusion due to crowding of  radio continuum features in the inner 
Galaxy.

Velocity profiles of H$_3^+$ and CO lines toward  the Galactic center indicate diffuse gas 
in the CMZ is expanding radially at a velocity of 150 \kms \citep{oka20}. This expanding radial   outflow could originate from 
the jet  interacting with molecular gas in the CMZ, as  described below. 
The geometry of the biconical 
outflow implies that  the locations of Sgr B2 and Sgr E are  
on the back and fronts sides of Sgr A* (see below). 
The jet picture quantified  below  can also be applied to molecular filaments  running parallel to 
the Galactic plane in Sgr E \citep{wallace22} (see Fig. 8). 
An alternative suggestion that has been put forth  is that 
stretching of molecular gas in Sgr E is due to 
its motion  along $x1$ orbits. The Sgr E complex lies at the intersection of  CMZ and the Galactic dust lane, so it is 
possible that 
the gravitational potential of the bar is  responsible for stretching  the Sgr E cloud into filaments \citep{wallace22}. 
This suggestion is based on limited ALMA observations over a small region showing CO filaments on a length scale of 2pc. 
It is not clear if this suggestion can  
 explain the  radial distribution of filament PAs pointing toward Sgr A*  on a scale of 300 pc. 

\subsection{A Model of an outflow from Sgr A* along the Galactic plane}

We consider a model in which Sgr A* is the source of an outflow that has sufficient ram pressure to 
distort and stretch thermal,  and nonthermal  materials that are  embedded within the outflow. In this picture, 
the outflow has  mass loss rate $\dot M$, velocity $u$, Lorentz factor 
$\gamma$ and is assumed to be directed into opposing cones, each with half-opening angle $\theta$ (see Figure 8 for a schematic diagram).   
The outflow 
delivers momentum at rate $\dot p = \dot M\gamma u$ into solid angle $\Omega = 8\pi\sin^2(\theta/2)$.  The ram pressure 
within an outflow cone at distance $d$ from Sgr A* is $\pram=\dot p/(\Omega d^2)$, i.e.
\begin{equation}
   \pram = \frac{\dot M\gamma u}{8\pi d^2\sin^2(\theta/2)}\,.
    \label{eqn:â€¢}
\end{equation}
Adopting nominal values $\dot M\gamma u/c = 
10^{-4}$\,M$_\odot$\,yr$^{-1}$, 
and $\theta = 20^\circ$ consistent with outflow parameters invoked on smaller scales \citep{zadeh12a,royster19,zadeh20},  we find 
that the ram pressure 
within the cones is $\pram/k = 2\times10^6$\,cm$^{-3}$\,K at  $d=300\,$pc. In order for the outflow to align nonthermal 
radio 
filaments with  the outflow direction, the ram pressure 
must  exceed the internal filament pressure. Assuming 
that the magnetic field B$\sim86\, \mu$G, the mass outflow rate  
has to exceed   
$10^{-4}$\,M$_\odot$\,yr$^{-1}$. 
 
There are also ionized and molecular  clouds,  such as the Sgr E complex, that are embedded within the outflow. 
To estimate the velocity gradient that this outflow might induce in an embedded or partially embedded cloud,  we note that 
cloud material with surface density $\Sigma$  experiences acceleration $a=\pram/\Sigma$.  From a standing start, the 
velocity acquired after being accelerated through distance $L$ is  $(2aL)^{1/2}$.  Clouds typically exhibit order-unity 
surface density variations, so the difference in the ram-pressure induced velocity across the cloud is of order $v$, and 
we write $\Delta  v \approx (2aL)^{1/2}$ (see Fig. 8).

We now apply this scenario to the Sgr A E cloud, which exhibits   $\Delta v \approx 20$\,km\,s$^{-1}$  over a scale of 0.5 
degrees \citep[see their Fig 10] {anderson20}, ie $L\approx70$\,pc at 8\,kpc.  
We use the outflow parameters adopted 
above,  $10^{-4}$\,M$_\odot$\,yr$^{-1}$.   
The mean integrated $^{13}$CO 2-1 line intensity is 17 K \kms \citep{anderson20}, and using 
$X_\mathrm{CO} = 10^{21}$ cm$^{-2}$  (K\,  km\, s$^{-1}$)$^{-1}$  \citep{schuller17},  we
obtain a hydrogen column $N_H = 1.7 \times 10^{22} \mathrm{cm}^{-2}$  and compute the clouds surface density using 
$\Sigma = 1.4 m_\mathrm{H} N_\mathrm{H}$. 
Then we find that the acceleration $a\approx 4\,\mathrm{pc\,Myr^{-2}}$, yielding $\Delta v\approx23$\,km\,s$^{-1}$, 
consistent with the observed velocity gradient.
The time scale needed to accelerate the material in this scenario is  $(2L/a)^{1/2}\approx 6\,$Myr,  places a lower limit 
on the age of the outflow.



One  advantage of this  model is that 
the large-scale jet-driven outflow is consistent with a picture that has independently
suggested to  explain multiple  observations at small scales from $\sim0.5$ to 25pc from  Sgr A* where a 
relativistic jet had previously  been 
proposed (see \cite{zadeh12a,royster19,zadeh20} and references therein).  

\section{Conclusion}

In summary, we have presented an analysis of the position angle distribution of the entire system of thermal and 
nonthermal radio filaments in the Galactic center. 
We use filtered MeerKAT images to  determine the position angle distributions of the population of 
nonthermal and 
thermal radio filaments. We noted alignment of filament directions within 20$^\circ$ of the Galactic plane that that 
suggested a scenario in which a collimated outflow from Sgr A* in the direction along the Galactic plane. 
The inferred jet is estimated to have a a mass flow rate of $10^{-4}$ \msol\, yr$^{-1}$  in order to 
align nonthermal  radio filaments and accelerate thermal gas along the Galactic plane.   This outflow 
emerges 
perpendicular to the equatorial plane of the accretion flow of Sgr A* and runs along the Galactic plane. 
The opening angle 
of the outflow is estimated to be $\sim40^\circ$ with respect to the Galactic plane where the relativistic jet lies, and 
the age of the outflow is at least 6 Myr.


\begin{acknowledgments}
Work by R.G.A. was supported by NASA under award number 80GSFC21M0002. The MeerKAT telescope is operated by the South African Radio Astronomy 
Observatory, which is a facility of the National Research Foundation, an agency of the Department of Science and Innovation. 
\end{acknowledgments}


\bibliographystyle{aasjournal}


\begin{figure}[ht!]
\epsscale{0.8}
\plotone{fig1a_LR.pdf}
\plotone{fig1b_PA_fig0_exclusions_v3_LR.pdf}
\caption{
{\it (a)} 
A mosaic MeerKAT image of the Galactic center at 20cm with a $4''$ resolution \citep{heywood22}. 
{\it  (b)} 
Similar to {\it (a)} but  filtered image with a $6.4''$ resolution \citep{zadeh22a}. 
The units  are reported in Jy 
beam$^{-1}$, where the beam area is that of the original 4$''$ beam. The filtering process reduces the numerical value of 
the brightness in 
filament  pixels by a factor of $\sim$0.13, with large variations due to the effect of the background removal. 
This factor is not applied to the filtered image intensities.
Supernova remnants and HII regions identified as  red circles  in both $(a)$ and $(b)$ are excluded in the analysis. 
}\end{figure}

\begin{figure}[ht!]
\plotone{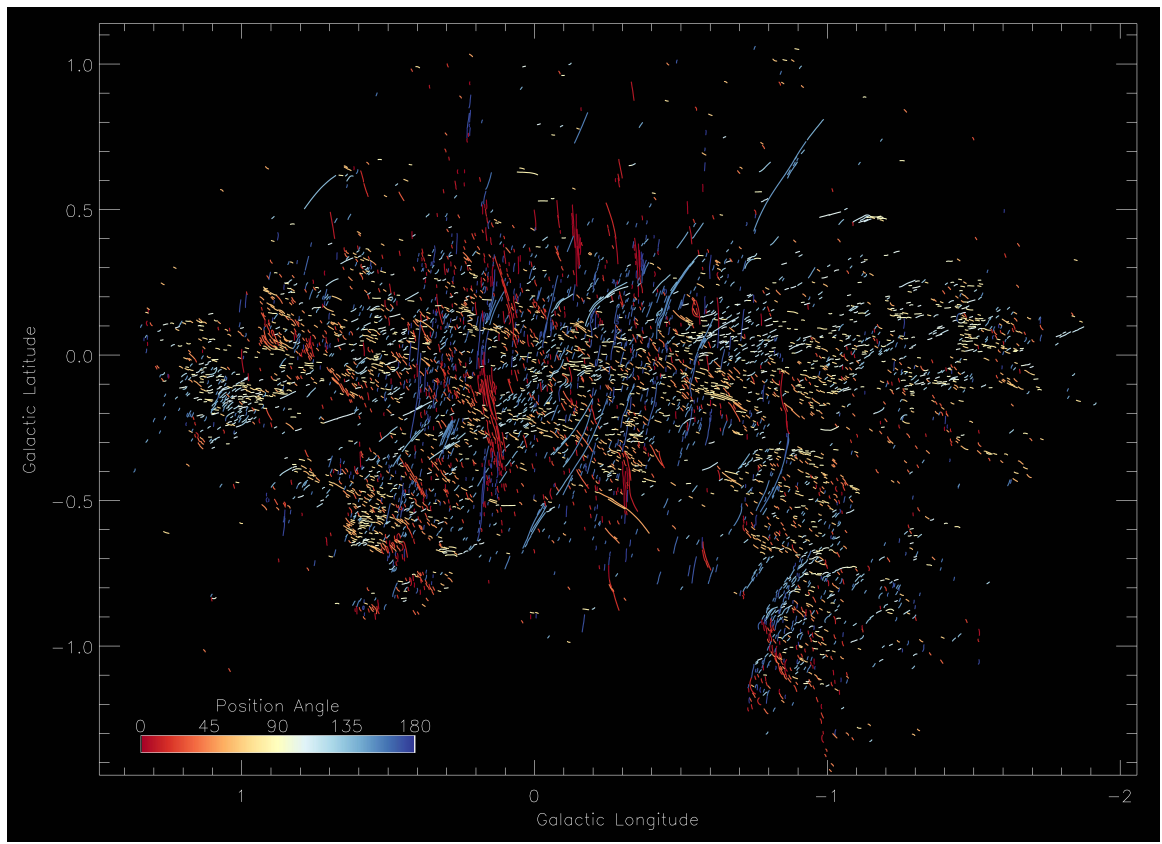}
\caption{
{\it (a)} 
Color-coded position angles for all identified short and long filaments  in the mosaic image (Fig. 1) are displayed.
(East of Galactic N is positive). 
}
\end{figure}

\addtocounter{figure}{-1}
\begin{figure}[ht!]
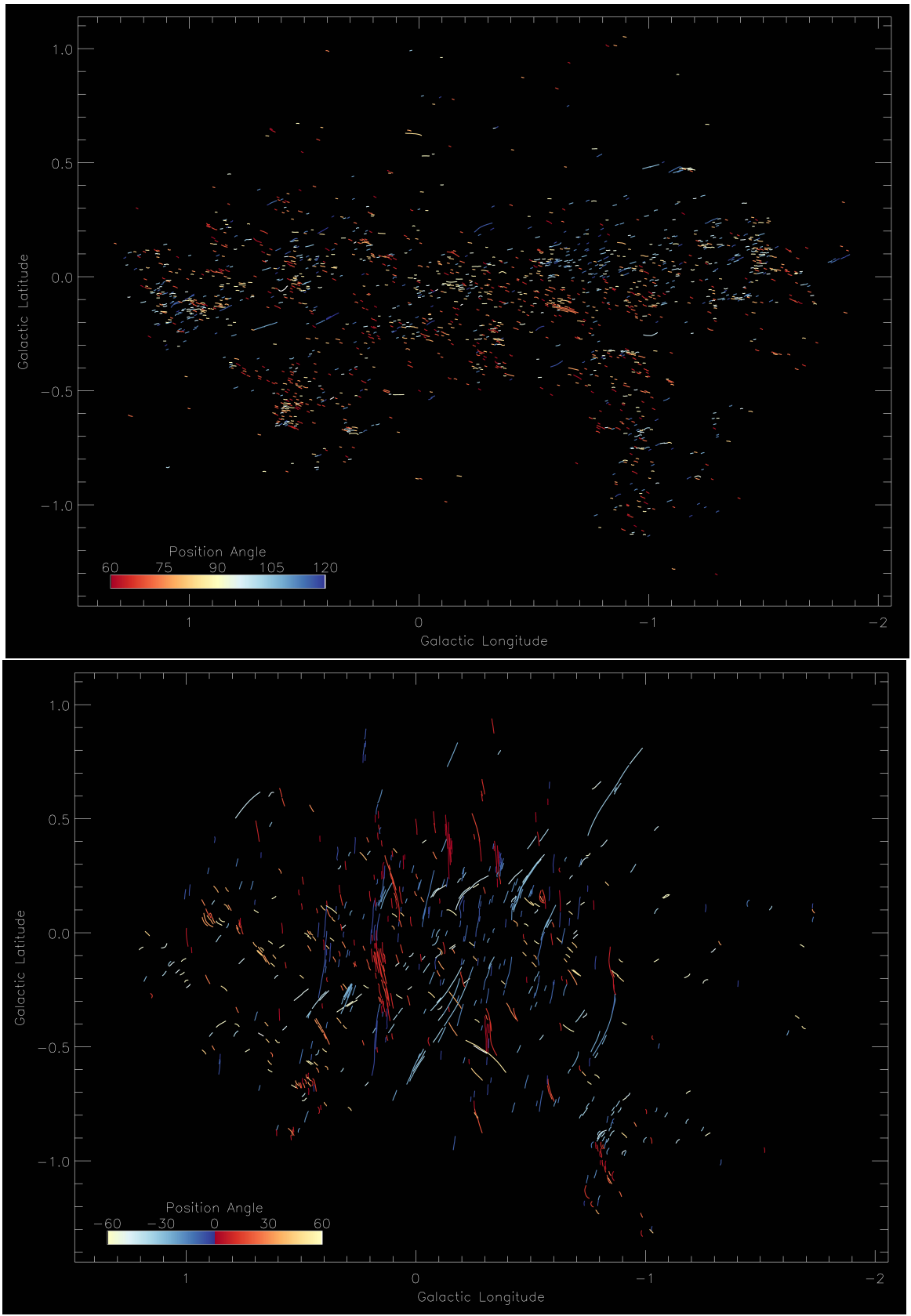

\plotone{fig2b_PA_v4.pdf}
\plotone{fig2c_PA_v4.pdf}
\caption{
{\it (b)} 
Similar to {\it (a)}  except that the color table is restricted, indicating a preferred direction of short filaments $L<66''$.  
The filament PAs run between $60^\circ$ and $120^\circ$. 
{\it (c)} 
Similar to {\it (b)}  except that the color table is restricted, selecting only  long filaments PAs $L>66''$. 
The filament PAs run between $-60^\circ$ and $60^\circ$. 
}
\end{figure}

\begin{figure}[ht!]
\plotone{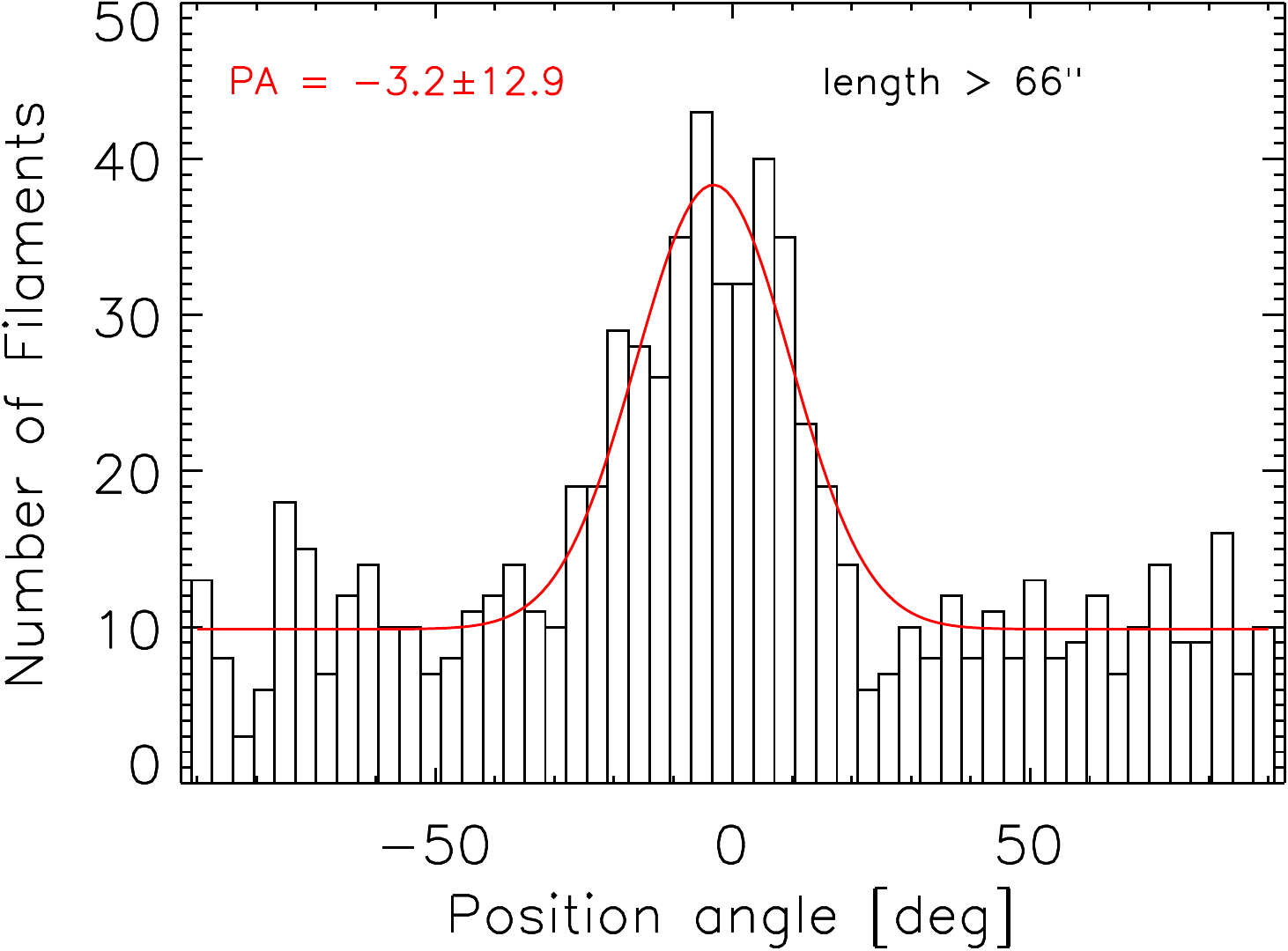}
\plotone{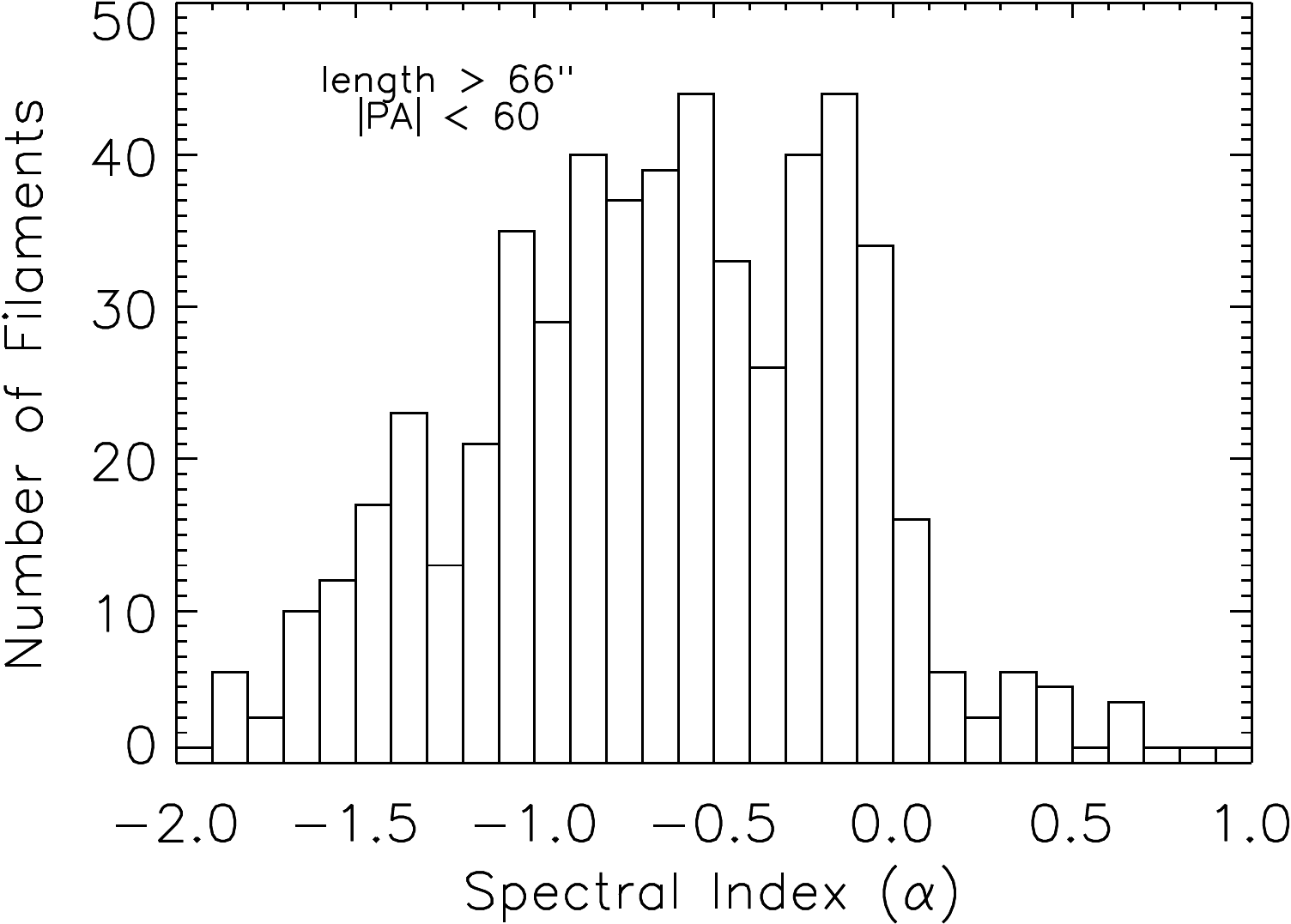}
\caption{
{\it (a)} 
A histogram of the PA distribution of filaments with lengths $L>66''$. 
Galactic north has PA$=0^\circ$. 
{\it (b)} 
Similar to  $(a)$ except for the spectral index distribution. Known SNRs and HII regions 
have been masked. 
\label{fig:general}
} 
\end{figure}

\begin{figure}[ht!]
\plotone{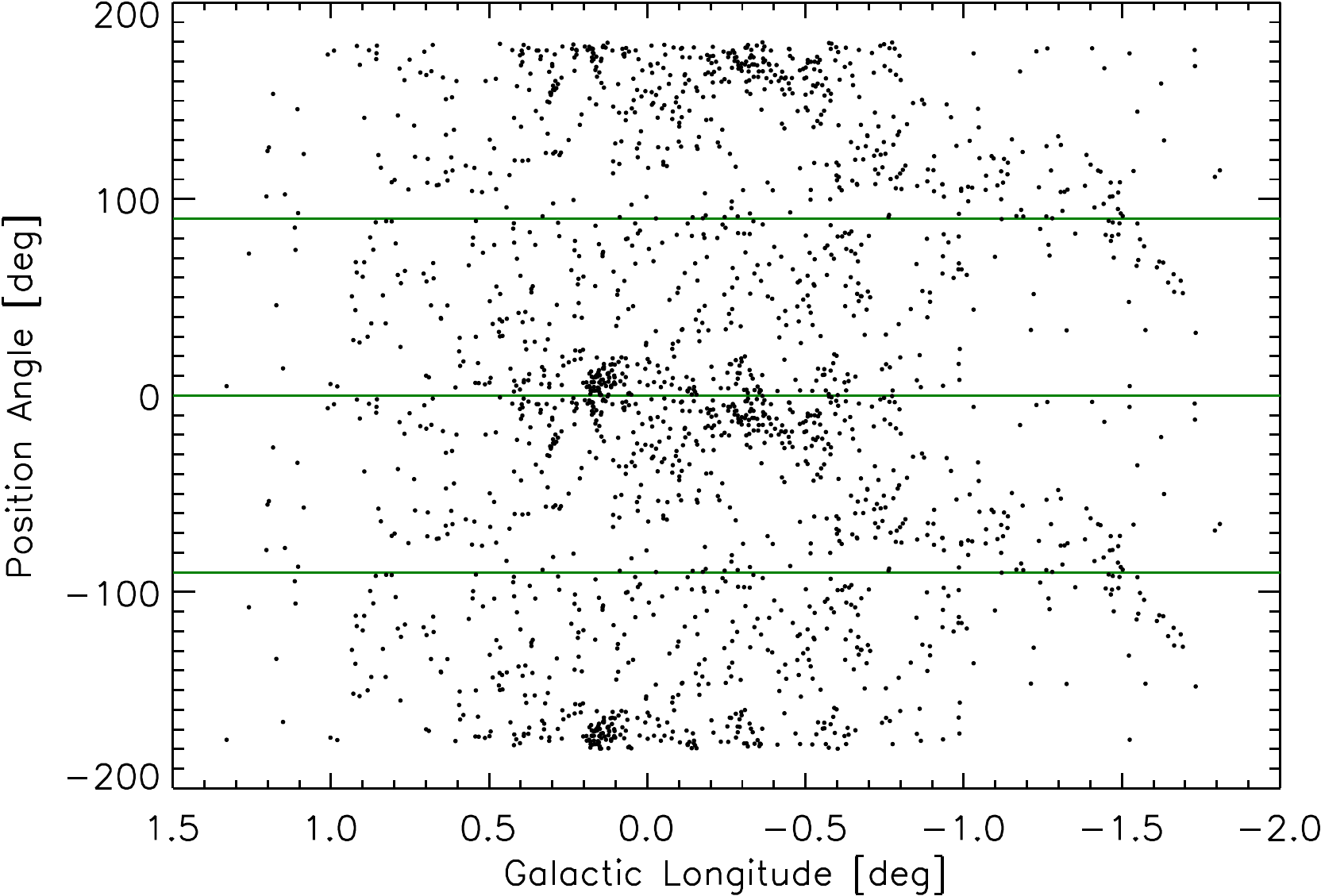}
\plotone{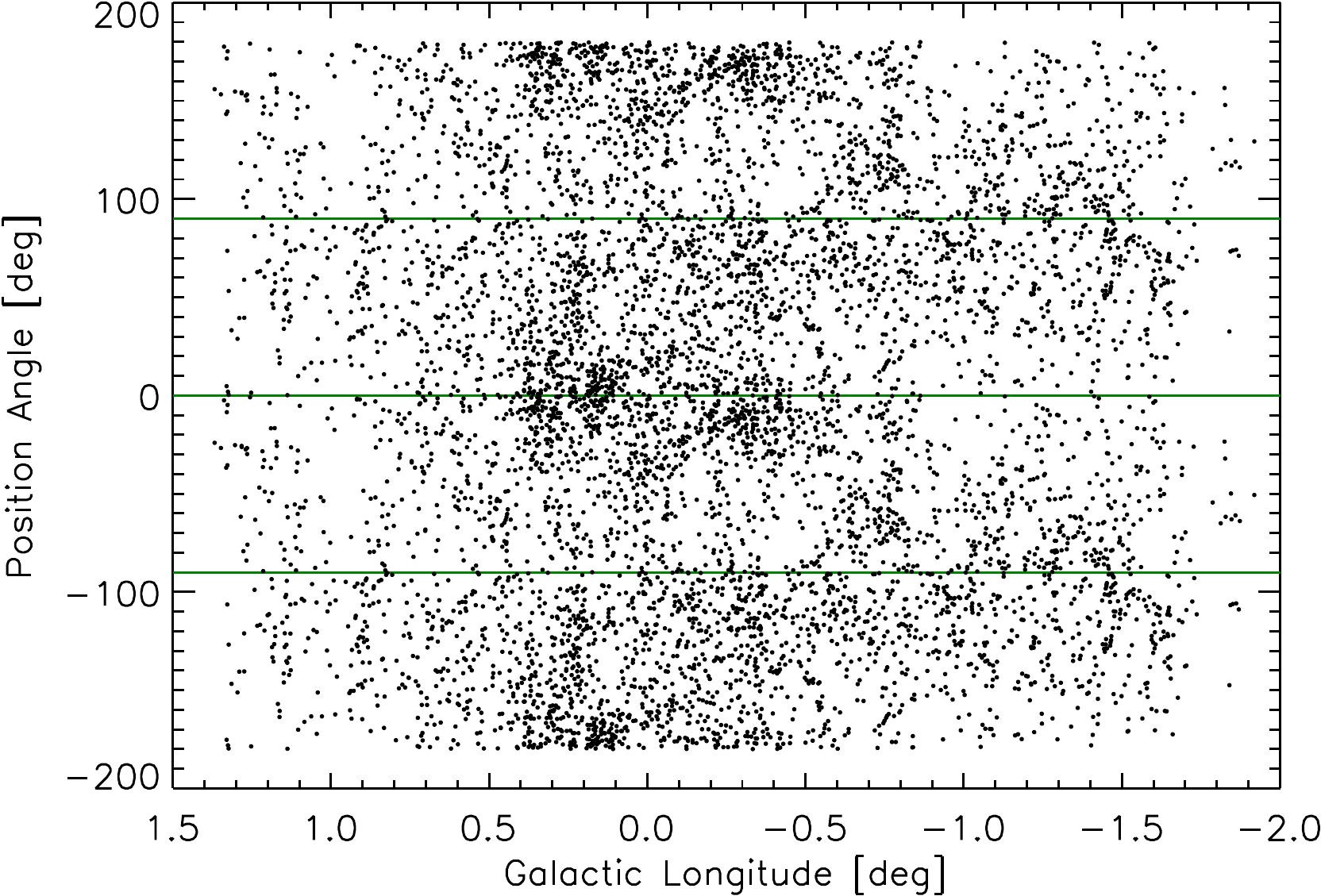}
\caption{
{\it (a)} 
The  PAs of filaments  with $L>66''$, represented by a point, is shown  as a function of  Galactic longitude. 
A  clustering of  PAs slightly greater and slightly less than $0^\circ$ are located near Galactic longitudes 
$l\sim0.2^\circ$ and  $\sim-0.3^\circ$, respectively.  
{\it (b)} 
Similar to (a) except that the filament PAs are restricted to  $L<66''$.
Excluded regions shown in Figure 1 are applied before (a) and (b)  were constructed. 
\label{fig:general}
} 
\end{figure}

\begin{figure}[ht!]
\plotone{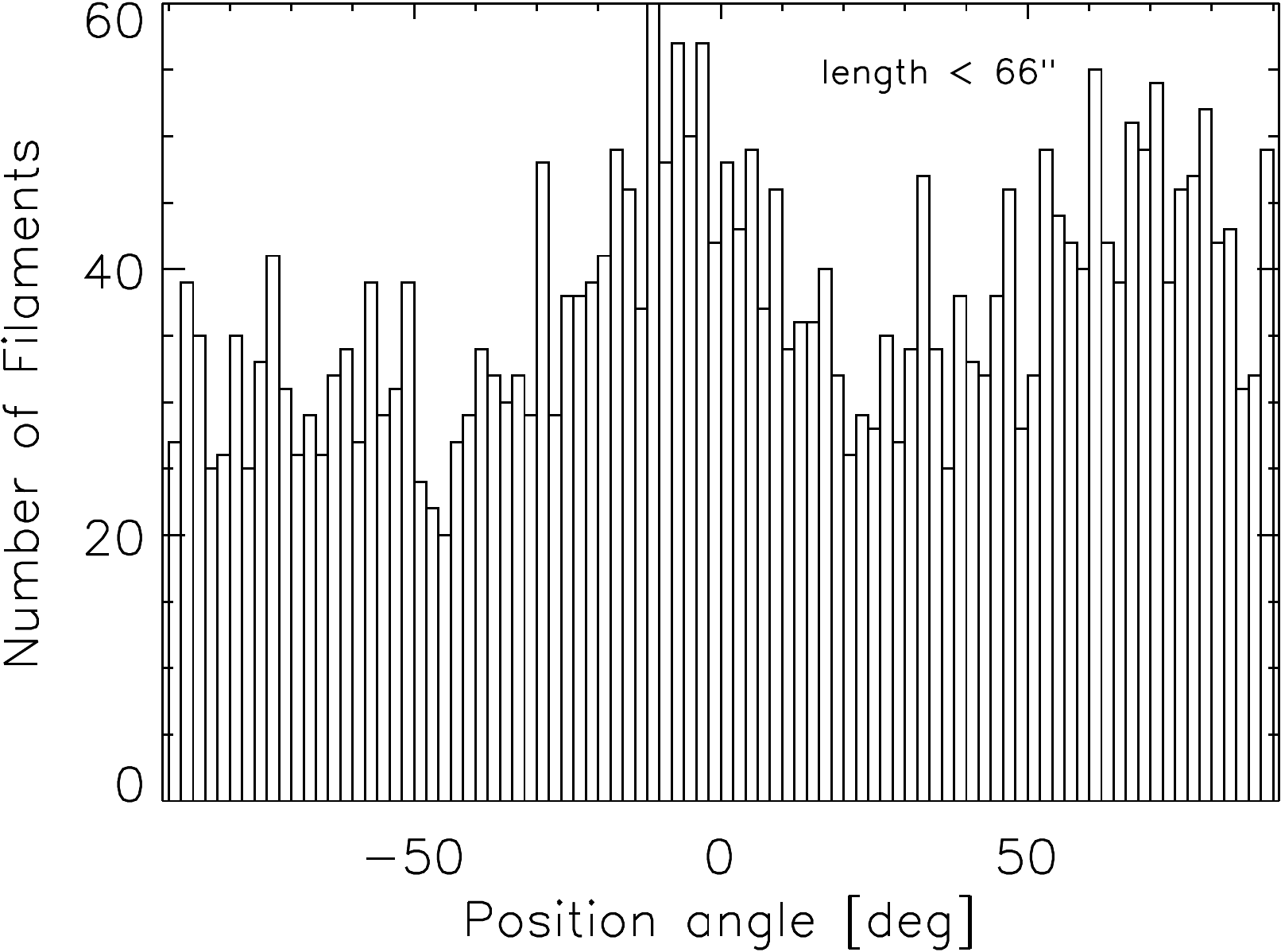}
\plotone{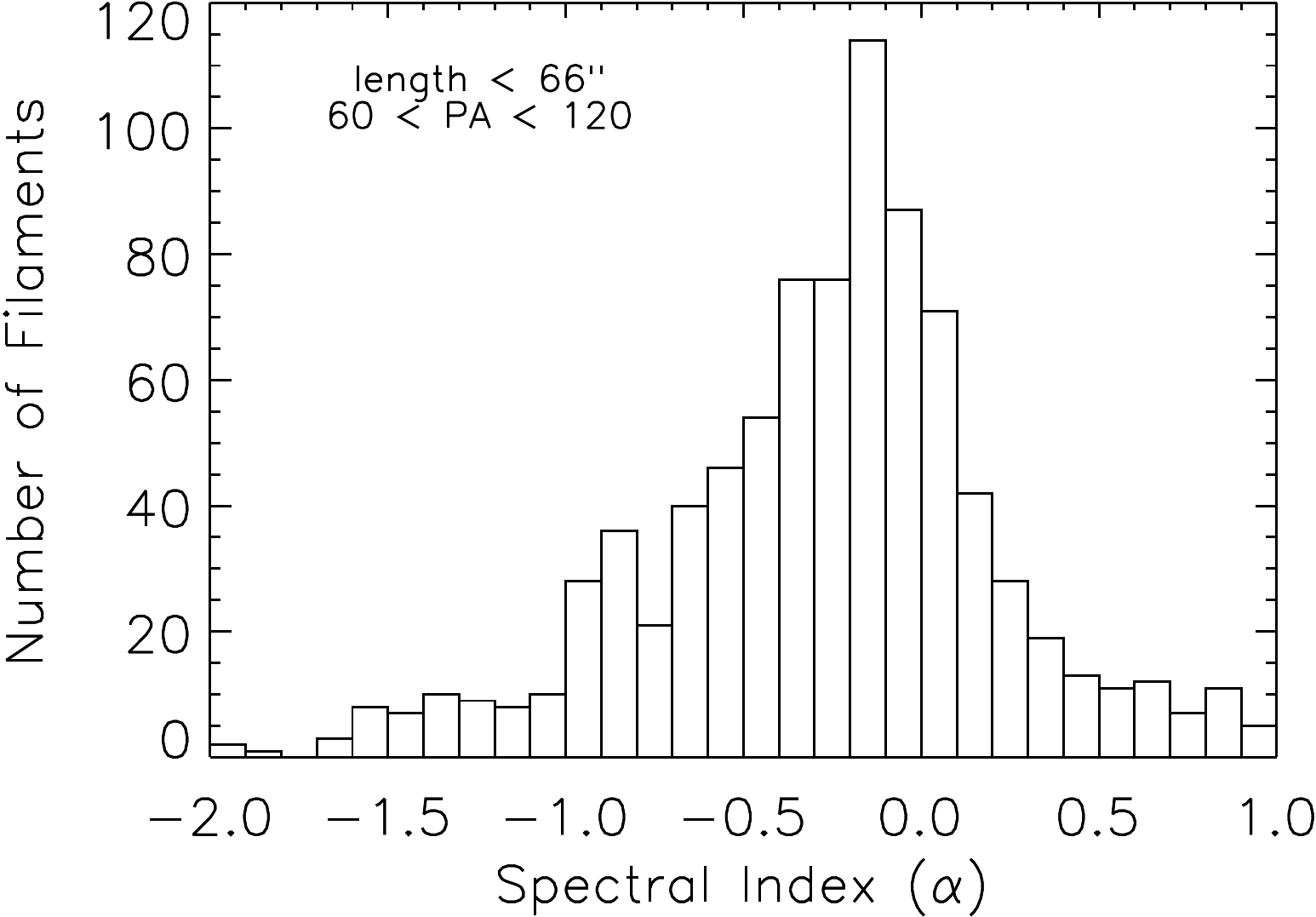}
\caption{
{\it (a)} 
Similar to Figure 3a except for short filaments ($L<66''$).
{\it (b)}
Similar to Figure 3b except for short filaments ($L<66''$).
\label{fig:general}
} 
\end{figure}

\begin{figure}[ht!]
\plotone{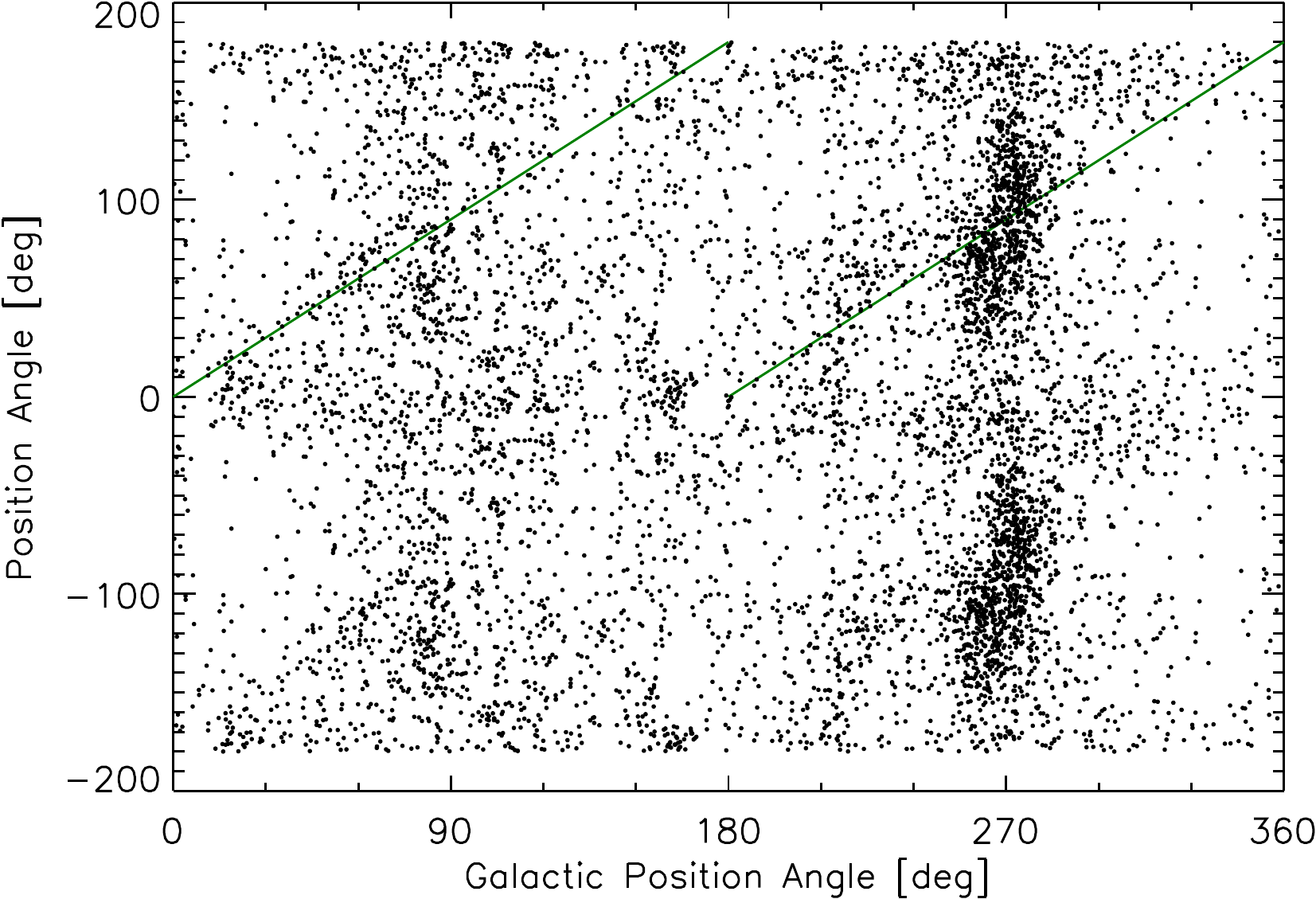}
\plotone{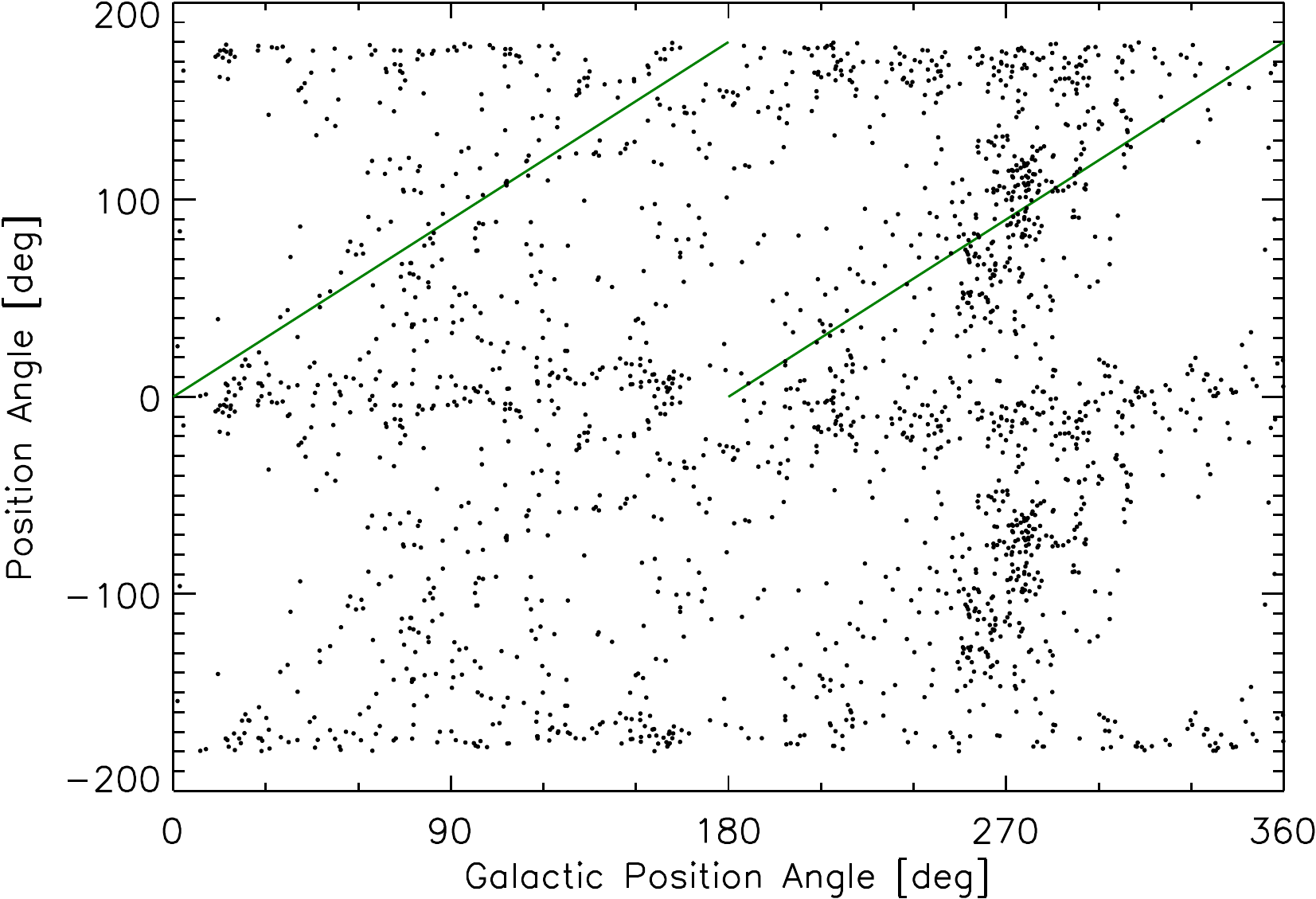}
\caption{
{\it  (a)} 
The clump of points near Galactic PA =270$^\circ$  with Galactic position angle GPA$= tan^{-1} l/b) < 270^\circ$  
identify shorter-filament PAs that are mostly 
below  the
radial trend (orange line). Those with 
GPA $> 270^\circ$  (4th quadrant) have PAs that are generally
above the radial trend with steeper values, suggesting that they are close to the Galactic plane. 
On the other hand, at GPA $< 270^\circ$  (3th quadrant) filament PAs  are generally
flatter, running away from the Galactic plane. 
{\it (b)} 
The same as (a) except for long filaments.  
\label{fig:general}} 
\end{figure}

\addtocounter{figure}{-1}
\begin{figure}[ht!]
\plotone{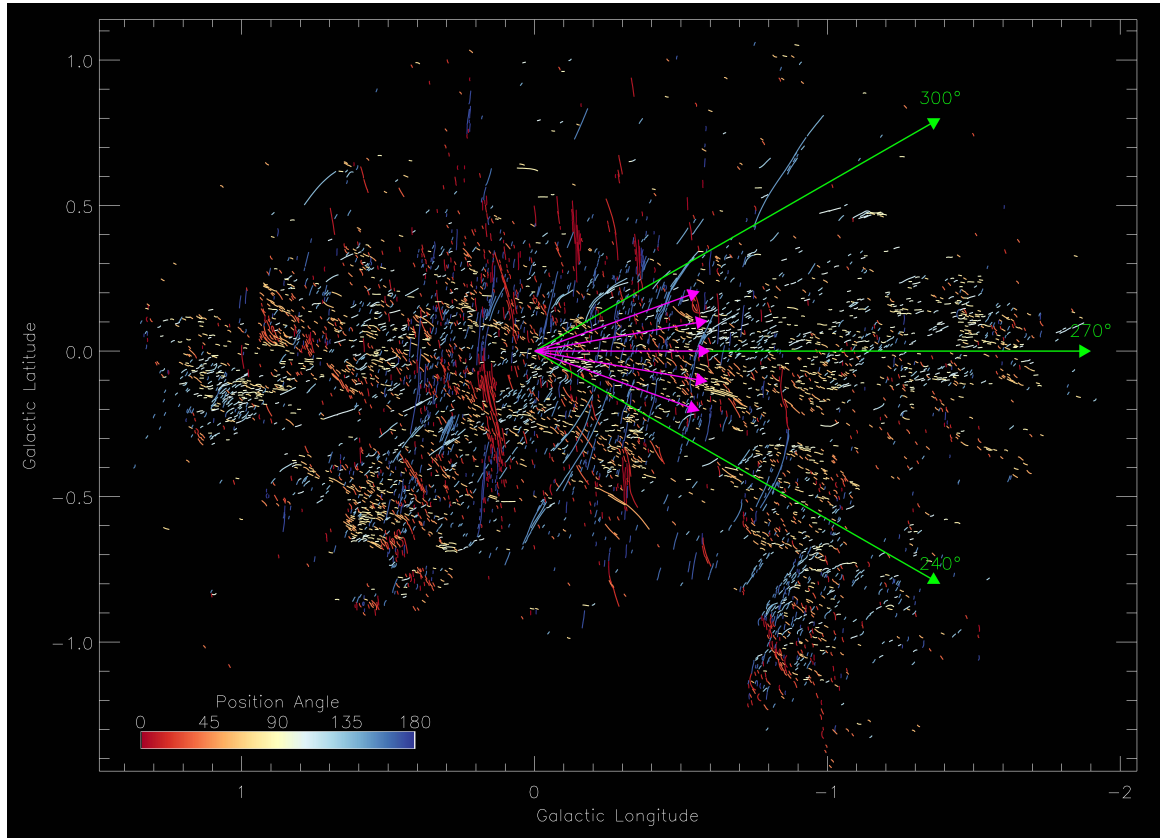}
\caption{
{\it  (c)} 
A diagram showing the PAs of short filaments within a cone pointed toward the direction of Sgr A*.
\label{fig:general}} 
\end{figure}

\begin{figure}[ht!]
\epsscale{0.8}
\plotone{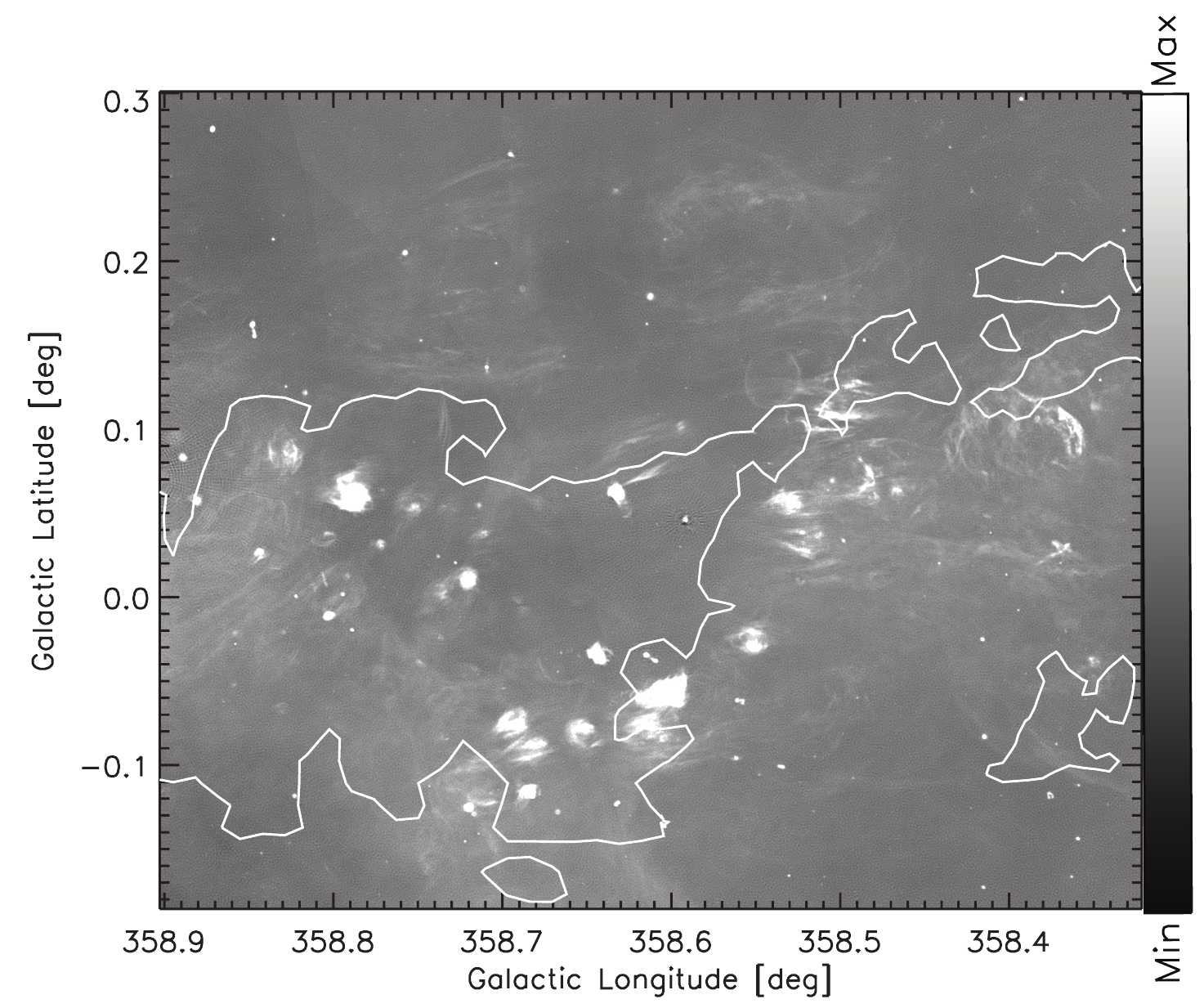}
\plotone{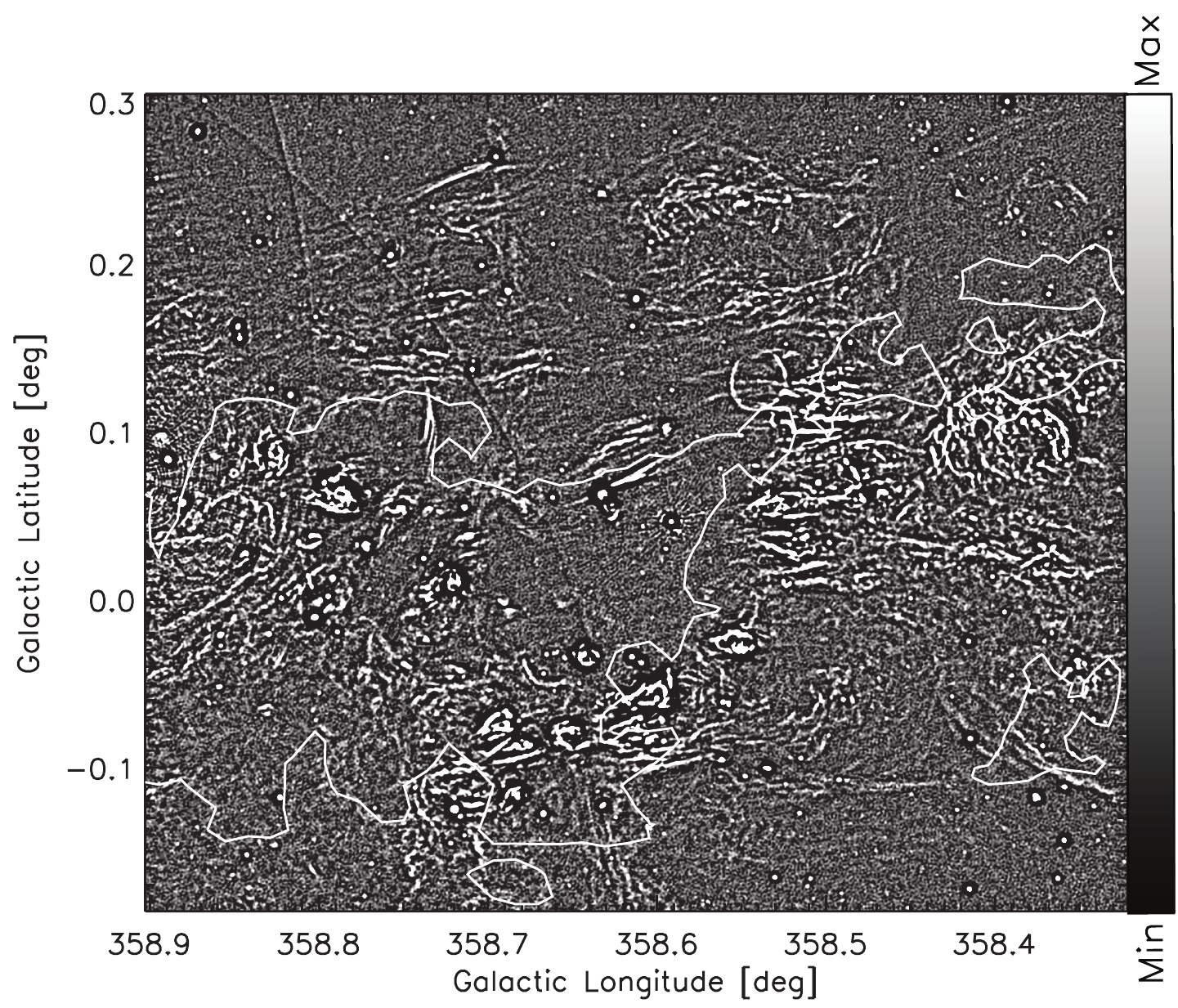}
\caption{
{\it  (a)}
Unfiltered continuum MeerKAT image of the Sgr E HII complex G358.7-0.0 at  20cm 
showing a cluster  of 
discrete HII regions surrounding a dark cavity coincident with the Sgr E molecular cloud 
\citep{heywood22}. 
The outline of $^{13}$CO (2-1)  line emission from the Sgr E cloud is  shown as white contour \citep{anderson20}.   
The range of grayscale is  [-5e-4,7e-4] Jy beam$^{-1}$. 
{\it (b)} 
Similar to (a) except filtered continuum image is displayed \citep{zadeh22a}.  
The range of grayscale is [-2e-5, 7e-5] Jy beam$^{-1}$. 
\label{fig:general}}  
\end{figure}

\addtocounter{figure}{-1}
\begin{figure}[ht!]
\plotone{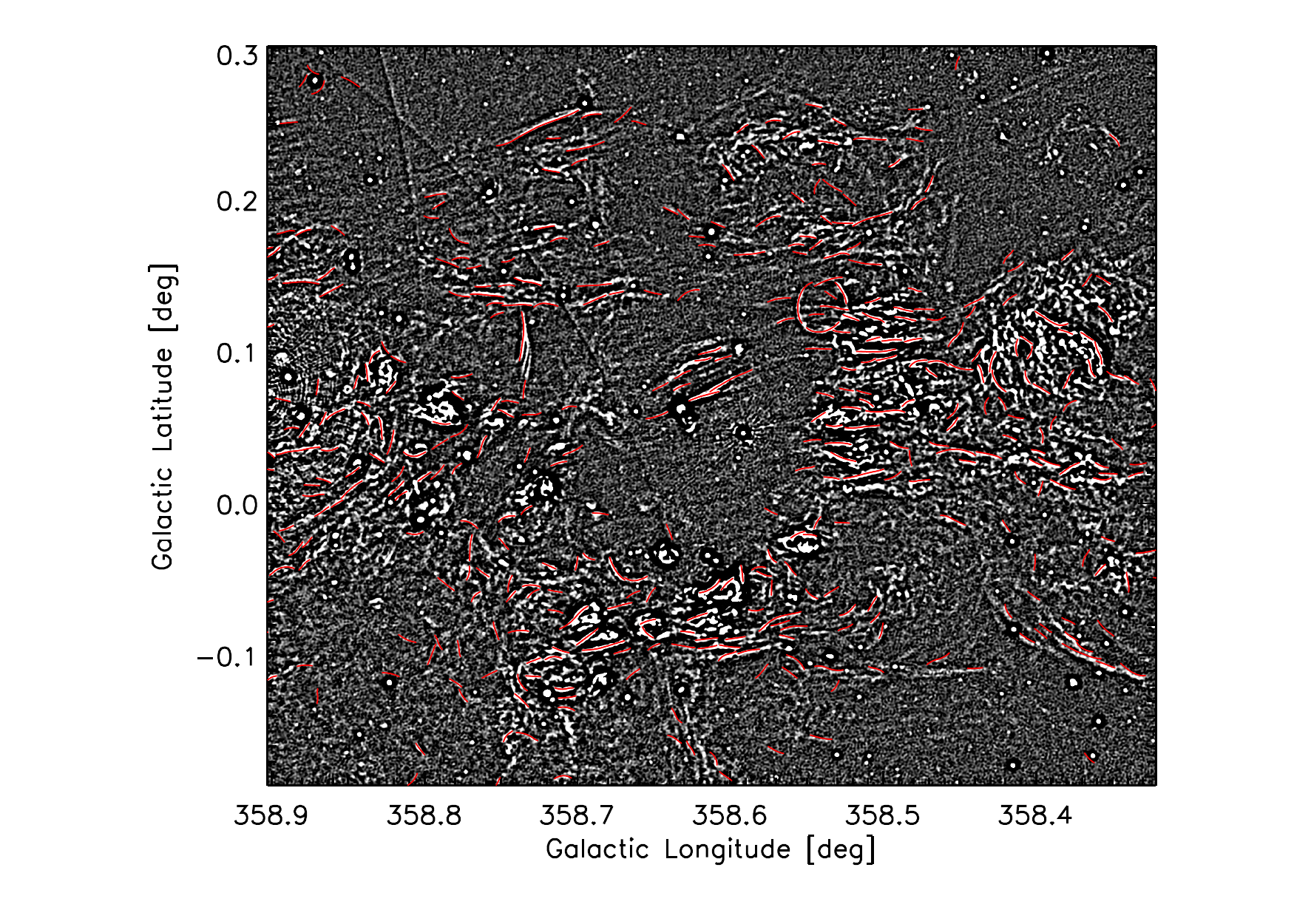}
\plotone{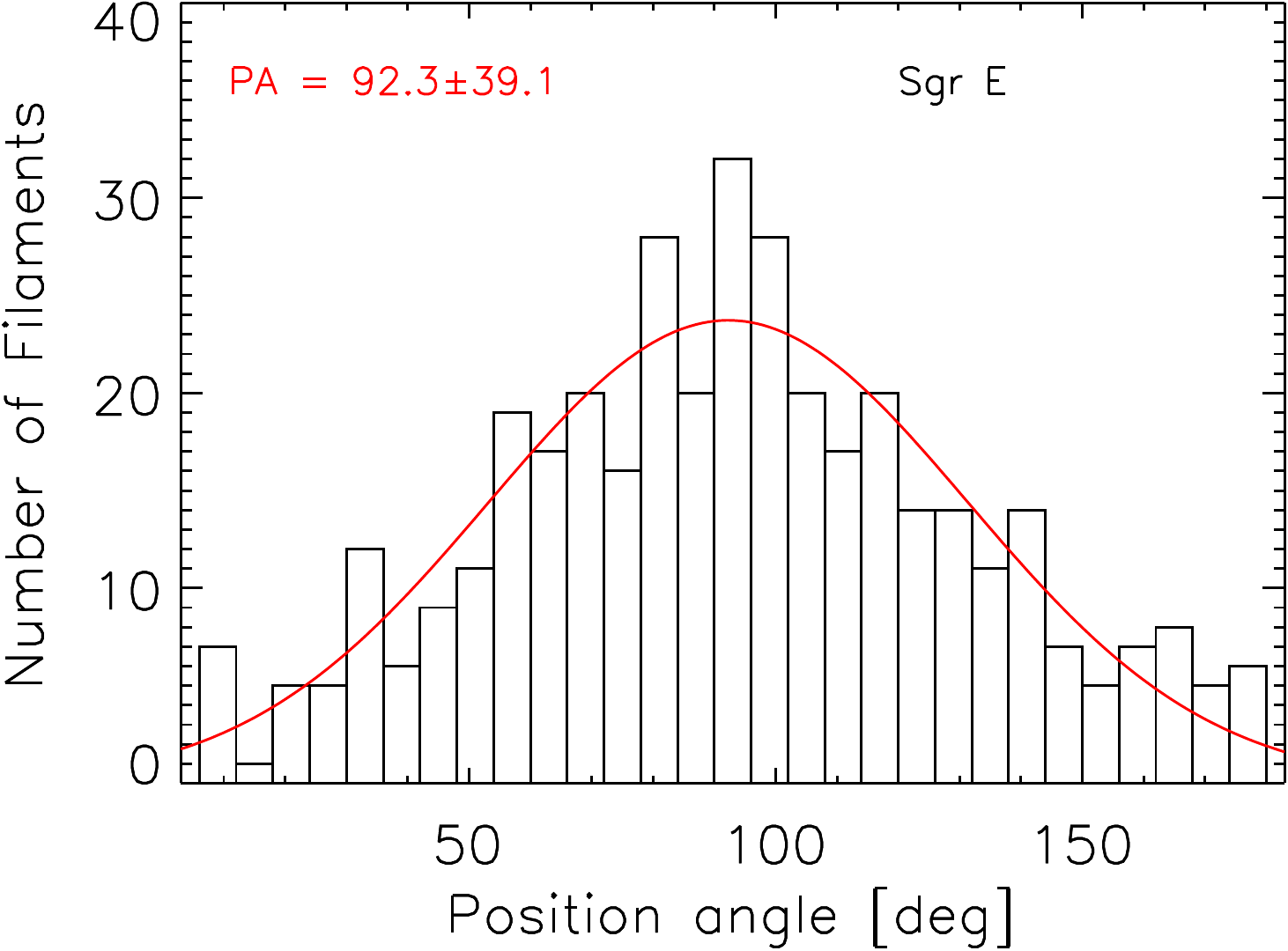}
\caption{
{\it  (c)} 
Same as (b) except that identified  filaments are  drawn in red. 
The range of grayscale [-2e-5, 7e-5] Jy beam$^{-1}$. 
{\it (d)} 
A histogram of filaments PAs in Sgr E  shows  that most filaments PA are distributed within 
$\sim40^\circ$. Red line shows a Gaussian fit to the histogram.   
of the Galactic plane. 
\label{fig:general}} 
\end{figure}

\begin{figure}[ht!]
\epsscale{0.9}
\plotone{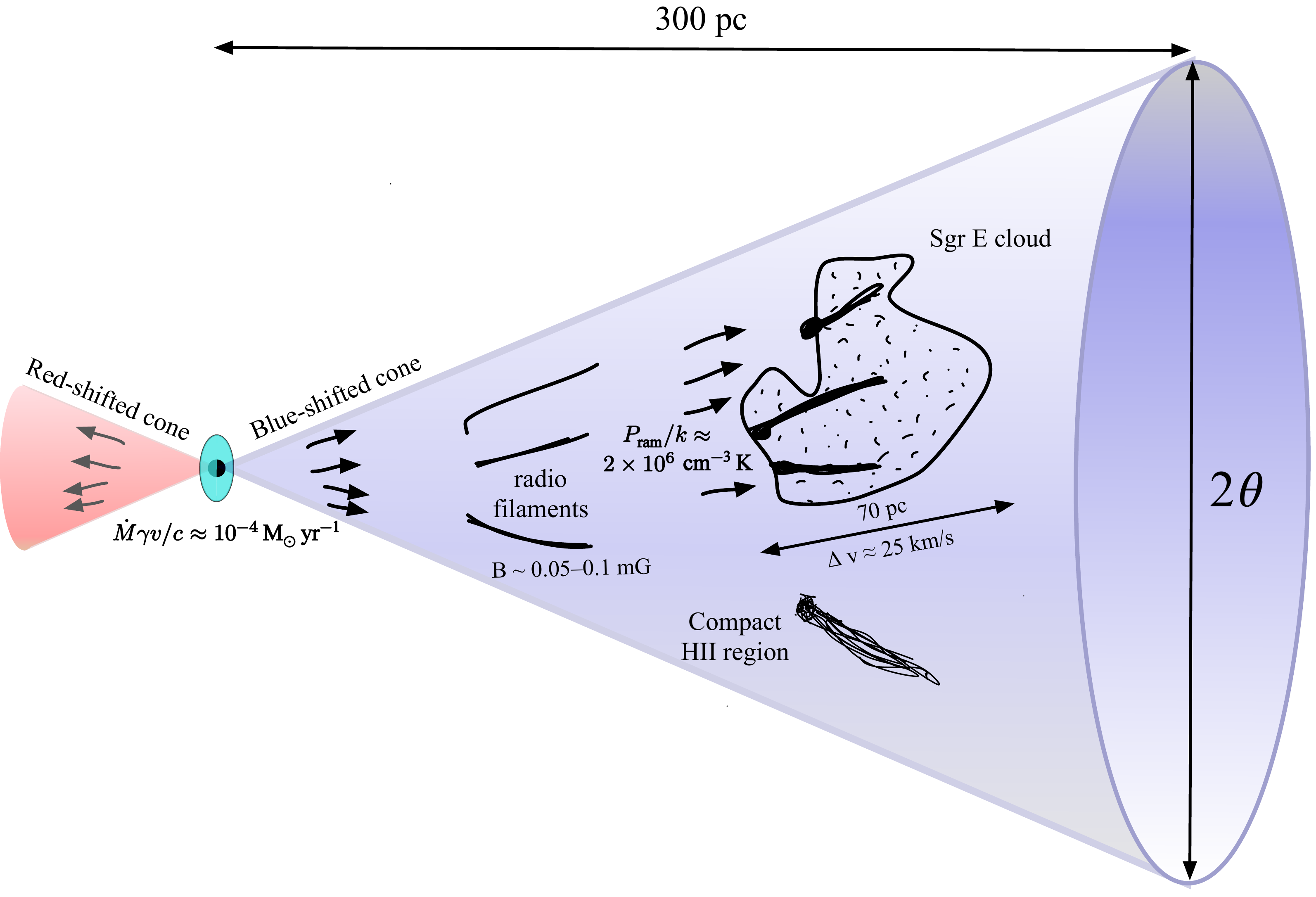}
\caption{
A schematic diagram of a jet-driven outflow with an opening angle  2$\theta$ along the Galactic plane. 
\label{fig:general}} 
\end{figure}

\end{document}